\documentclass[aps,prl,superscriptaddress,twocolumn]{revtex4-1}

\usepackage{graphicx}
\usepackage{amsmath}
\usepackage{amssymb}
\usepackage{amscd}
\usepackage{bm}
\usepackage{enumerate}
\usepackage{type1cm}
\usepackage{lettrine}
\usepackage{mathrsfs}
\usepackage{calrsfs}
\usepackage{epsfig}
\usepackage{subfigure}
\usepackage{psfrag}
\usepackage{color}

\def\beq{\begin{equation}}
\def\eeq{\end{equation}}
\def\bea{\begin{eqnarray}}
\def\eea{\end{eqnarray}}

\makeatletter
\renewcommand*{\@fnsymbol}[1]{\ensuremath{\ifcase#1\or \dagger\or *\or \ddagger\or
   \mathsection\or \mathparagraph\or \|\or **\or \dagger\dagger
   \or \ddagger\ddagger \else\@ctrerr\fi}}
\makeatother

\begin{document}

\title{Shape dynamics of growing cell walls}
\author{Shiladitya Banerjee}
\affiliation{James Franck Institute, The University of Chicago, Chicago IL 60637}
\author{Norbert F. Scherer}
\thanks{To whom correspondence may be addressed. Email: dinner@uchicago.edu or nfschere@uchicago.edu}
\affiliation{James Franck Institute, The University of Chicago, Chicago IL 60637}
\affiliation{Institute for Biophysical Dynamics, The University of Chicago, Chicago IL 60637}
\affiliation{Department of Chemistry, The University of Chicago, Chicago IL 60637}
\author{Aaron R. Dinner}
\thanks{To whom correspondence may be addressed. Email: dinner@uchicago.edu or nfschere@uchicago.edu}
\affiliation{James Franck Institute, The University of Chicago, Chicago IL 60637}
\affiliation{Institute for Biophysical Dynamics, The University of Chicago, Chicago IL 60637}
\affiliation{Department of Chemistry, The University of Chicago, Chicago IL 60637}

\begin{abstract}
We introduce a general theoretical framework to study the shape dynamics of actively growing and remodeling surfaces. Using this framework we develop a physical model for growing bacterial cell walls and study the interplay of cell shape with the dynamics of growth and constriction. The model allows us to derive constraints on cell wall mechanical energy based on the observed dynamics of cell shape. We predict that exponential growth in cell size requires a constant amount of cell wall energy to be dissipated per unit volume. We use the model to understand and contrast growth in bacteria with different shapes such as spherical, ellipsoidal, cylindrical and toroidal morphologies. Coupling growth to cell wall constriction, we predict a discontinuous shape transformation, from partial constriction to cell division, as a function of the chemical potential driving cell wall synthesis. Our model for cell wall energy and shape dynamics relates growth kinetics with cell geometry, and provides a unified framework to describe the interplay between shape, growth and division in bacterial cells.
\end{abstract}

\maketitle
\section{Introduction}
Understanding the growth of structures in living systems presents new challenges for soft matter theory owing to the interplay of irreversible dynamics and mechanochemical forces.  In turn, elucidating how intrinsic molecular factors and extrinsic environmental factors combine quantitatively to determine morphologies is important for understanding many biological processes, including wound healing, tissue morphogenesis, tumor metastasis, and plant cell wall formation. The purpose of this paper is to report a theoretical framework for modeling the dynamics of actively growing and remodeling shapes. In particular, we investigate the growth of bacterial cell walls, which epitomize growing active matter. Active growth arises from cell wall enzymes that catalyze the assembly reactions driving the synthesis of cell wall material.

Bacteria exhibit a remarkable diversity in cell shapes and sizes~\cite{pinho2013}. The shapes of most bacteria are defined by the peptidoglycan cell wall, in association with cytoskeletal proteins and internal turgor pressure. Cell walls are thicker and stiffer than most polymeric membranes and are capable of maintaining cell shapes while sustaining large amounts of turgor pressure. The significant variety of shapes ranging from spherical cocci to rod-shaped \textit{E.\ coli} to crescent-shaped \textit{C. crescentus} implies that the maintenance of each specific shape requires a distinct physical mechanism\cite{cabeen2010,gahlmann2014}. 

Cell shape has a direct relation to the observed quantitative laws governing cell size growth. For instance, exponential longitudinal growth is observed in rod-like bacterial cells, such as \textit{E.\ coli}~\cite{wang2010}, and \textit{B. subtilis}~\cite{koch1993}, or crescent-shaped \textit{C.\ crescentus}~\cite{iyer-biswas2014}, and even in eukaryotic cells such as the ellipsoidal shaped budding yeast~\cite{di2007}. In addition, recent experiments on \textit{S. aureus}, a model system for round bacteria, reveal that the cell volume grows exponentially throughout the cell cycle with increasing aspect ratio~\cite{zhou2015}. Thus, an anisotropic cell geometry can be linked to their exponential growth via lateral peptidoglycan insertion. 

Starting with Koch's hypothesis that surface stresses determine bacterial cell shape~\cite{koch2001}, a number of theoretical models have been proposed in recent years to account for the shape and growth of bacterial cell walls. These include growth by plastic deformations as surface stresses exceed a critical value~\cite{boudaoud2003}, elastic growth of peptidoglycan networks driven by assembly reactions~\cite{jiang2010}, and dislocation driven growth of partially ordered peptidoglycan structures~\cite{amir2012}. These models however do not make clear the relationship between cell shape, kinetics of growth and constriction, and the mechanochemical energies driving growth. For instance, how does exponential longitudinal growth arise from isotropic pressure in rod-like cells? How does cell shape influence growth and division kinetics? 

Motivated by our recent experimental work that provides detailed growth and contour data of single \textit{C.\ crescentus} cells across a large number of generations~\cite{iyer-biswas2014,wright2014}, we introduce a general theoretical model for the shape dynamics of growing cell walls based on a principle of minimal energy dissipation. For a bacterial cell wall, the dissipative forces arise from the insertion of peptidoglycan strands, whereas the driving forces arise from changes in the mechanochemical energy, $E$, associated with maintaining the shape of the cell wall. The dependence of $E$ on cell geometry directly determines which shape parameters grow and which are size limited. We discuss how the condition of growth, more specifically exponential growth, imposes constraints on the form of the energy function. We show that our model for cell shape dynamics encompasses previous phenomenological models of cell wall growth for specific geometries such as cylinders or spheres~\cite{jiang2011,jiang2012}. 

We use the model to study the interplay of cell shape, growth and division control in bacteria. With new shape analysis of our experimental data, we demonstrate how cell shape features such as width and curvature can influence the rate of cell size growth. In addition to elucidating the relationship between cell shape and growth dynamics, our model is capable of describing cell wall constriction. We predict that a threshold chemical potential for septal synthesis is required for completing cell division such that on reaching the threshold, cell shape discontinuously switches from partial to full constriction. We compare the predictions of our model with available experimental data on cell shape and growth of single bacterial cells.

\section{Theoretical Framework}

\subsection{Equations of Shape Dynamics} 
\begin{figure}
\centering
\includegraphics[width=\columnwidth]{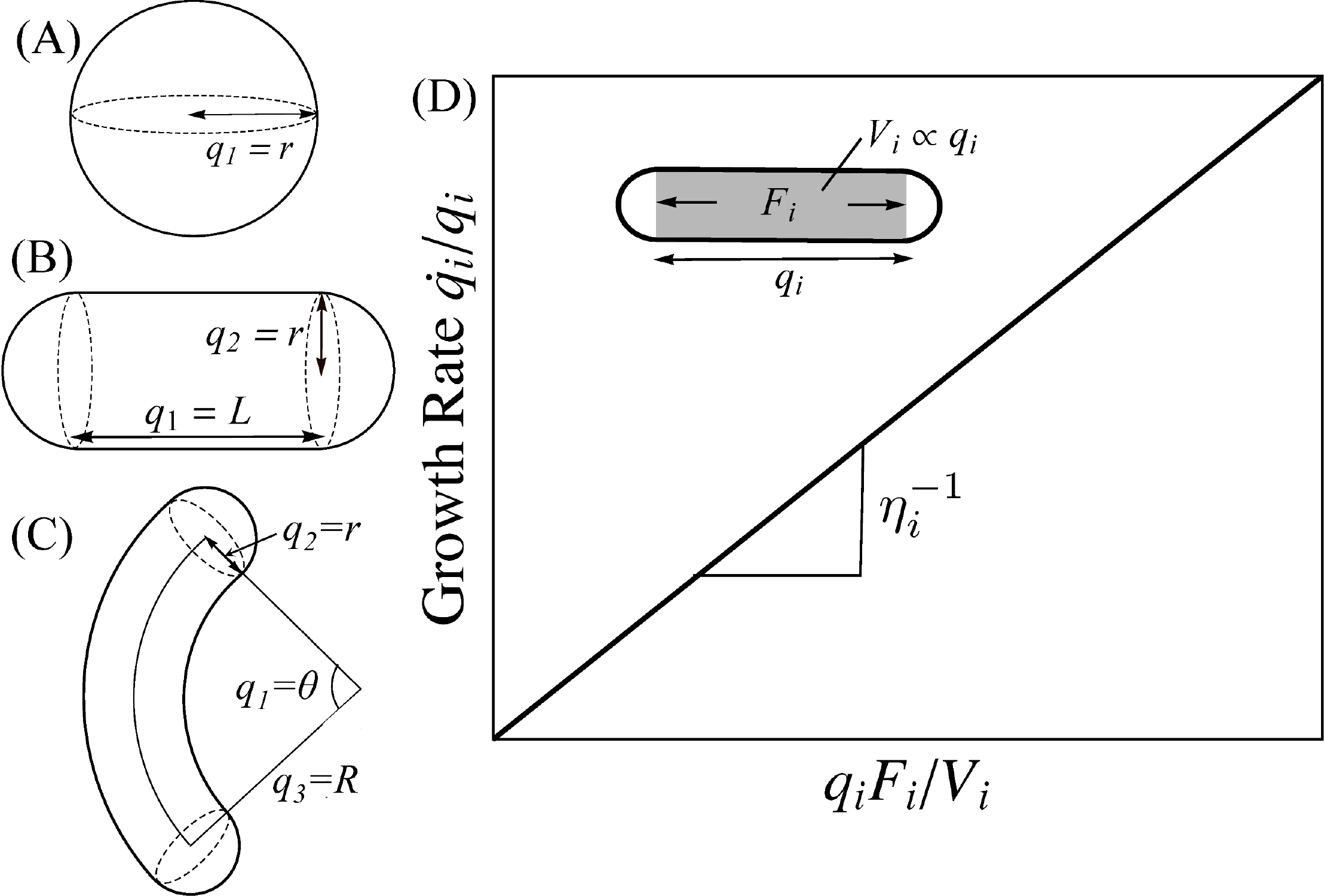}
\caption{Examples of shape parameters for (A) spherical, (B) cylindrical, and (C) curved cells. (D) Illustration of the growth law. Rate of growth of the shape parameter $q_i$ is proportional to the energy dissipated ($q_iF_i$) per unit volume ($V_i$). Inset: Physical picture of growth for a rod-like cell where the shaded region represents the dissipated volume and $F_i$ is the driving force.}
\label{fig:schematic}
\end{figure}

We parametrize the geometry of the cell wall by $N$ shape degrees of freedom specified by the generalized coordinates $q_i$ ($i=1,\ldots,N$) and the generalized velocities $\dot{q}_i$. For example, a sphere is parametrized by its radius, whereas a cylinder has two degrees of freedom, its radius and length (Fig.~\ref{fig:schematic}A-C). For a general surface described by a mesh of triangles, the generalized coordinates are given by the individual vertex positions. 
The mechanical energy of the cell wall is a function of the generalized coordinates, $E^{m}(\{q_i\})$. The generalized forces driving changes in cell wall geometry are given by the derivatives of the energy function, $F_i^{m}=-\partial E^{m}/\partial q_i$, $\forall i$. The equilibrium shape of the cell wall is simply given by minimizing the mechanical energy, $\partial E^{m}/\partial q_i=0$, $\forall i$.

In the absence of external forces and strong thermal noise, the mechanical forces ${\bf F}^{m}$ counterbalance the active forces (${\bf F}^a$) and the dissipative forces (${\bf F}^d$) associated with irreversible cell wall growth, ${\bf F}^a + {\bf F}^d+{\bf F}^{m}=0$. The active forces are of non-equilibrium origin and they arise from distributed molecular components in the cell that convert chemical energy into mechanical work.  For example, in the case of a growing bacterial cell wall, the active forces arise from cell wall enzymes that catalyze the assembly reactions driving peptidoglycan synthesis. We define the energy due to active processes as $E^a=\sum_i \int dq_i F_i^a$.

The dissipative forces are not symmetric under time-reversal and they are derived from minimizing the rate of energy dissipation, $\mathcal{D}$, using $F^d_i=-\partial \mathcal{D}/\partial \dot{q}_i$. The dissipation function, $\mathcal{D}$, represents the amount of work done to the medium when the shape deforms at a rate $\dot{q}_i$. The force-balance relation follows from minimizing the total rate of energy change in the system and the medium, i.e., $\partial(\dot{E}^m + \dot{E}^a+\mathcal{D})/\partial\dot{q}_i=0$, which is a statement of Rayleigh's principle of least energy dissipation~\cite{rayleigh1873}. This variational principle can be shown to hold for general irreversible processes and is equivalent to maximizing the rate of the entropy production in the system~\cite{doi2011}. 

The rate of dissipated energy is given by $\mathcal{D}=\frac{1}{2}\sum_i V_i \sigma_i (\dot{q}_i/q_i)$, where $V_i$ is the volume over which dissipation of $q_i$ occurs, $\sigma_i$ is the dissipative stress, and $\dot{q}_i/q_i$ is the strain rate. Assuming the deformation is small compared to current cell size, the dissipative stress is given by $\sigma_i=\eta_i (\dot{q}_i/q_i)$, where $\eta_i$ is the associated viscosity constant. The resultant equations of motion are
\begin{equation}\label{eqmotion1}
\eta_i V_i \frac{\dot{q}_i}{q_i^2}=-\frac{\partial E^m}{\partial q_i} + F_i^a=F_i\;, \forall i\;,
\end{equation}
such that the rate of growth is proportional to the total energy dissipated per unit volume ($F_i q_i/V_i$), as illustrated in Fig.~\ref{fig:schematic}D. In the limit of a rod-like cell, when the dissipated volume scales linearly with cell length, $V_i \propto q_i$ (Fig.~\ref{fig:schematic}D, inset), our model reduces to the phenomenological growth model proposed by Jiang and Sun~\cite{jiang2012}.
Eqn~\eqref{eqmotion1} can be written in a simple and instructive form by choosing logarithmic strain, $\Phi_i(t)\equiv\log{\left[q_i(t)/q_i(0)\right]}$, as our new dynamic variable,
\begin{equation}\label{eq:shapedyn}
\eta_i \frac{d\Phi_i}{dt}=-\frac{1}{V_i}\frac{\partial E}{\partial \Phi_i}\;, \forall i\;,
\end{equation}
where we defined $E=E^{m}+ E^{a}$, as the total internal energy of the system. Eqn~\eqref{eq:shapedyn} represents the familiar constitutive law of Newtonian flow such that the internal stress (right hand side), is proportional to the rate of strain. Eqn\ \eqref{eq:shapedyn} also illustrates that the laws of shape dynamics are isomorphic to the overdamped motion of a particle with coordinate $\Phi_i(t)$ in a potential $E$. 

\subsection{Energy requirements for exponential growth}

We now discuss how the specific form of the growth law puts constraints on the scaling of the energy function with the shape parameters. A necessary condition for growth in the shape parameter $q_i$ is $\partial E/\partial q_i<0$. Growth is arrested when $\partial E/\partial q_i=0$. It follows from eqn~\eqref{eqmotion1} that $q_i(t)$ grows exponentially if $E$ scales as the dissipated volume $V_i$, such that a constant stress $E/V_i$ drives material growth. For a growing bacterial cell wall, we assume that dissipation is dominated by peptidoglycan insertion over the thin shell defining the cell wall, such that $V_i=hA_i$, where $A_i$ is the surface area over which dissipation occurs and $h$ is the thickness of the cell wall assumed to be constant and uniform. A minimal model for exponential growth thus requires $E\propto -A_i$.

For a thin spherical shell of radius $r$ the dissipative volume scales as $V\propto r^2$. It then follows from the dynamics of cell radius, $\dot{r}\propto -\partial E/\partial r$, that for cells to grow exponentially i.e., $\dot{r}\propto r$, the energy would need to scale as $E\propto -r^2$. Using eqn~\eqref{eqmotion1}, we can thus conclude that a minimal energy model for exponential growth of isotropic cells is given by $E=-\varepsilon A$, where $A$ is the surface area and $\varepsilon$ is a positive constant representing the chemical potential for adding unit surface area.

As an example of growth dynamics in anisotropic cells, we consider a thin cylindrical shell of length $L$ and radius $r$. From eqn~\eqref{eqmotion1} it follows that the radius and the length grow exponentially with rates, $\dot{L}/L\propto -r^{-1}\partial E/\partial L$, and $\dot{r}/r\propto -L^{-1}\partial E/\partial r$. The shape-dependence of growth rates implies that a minimal energy model $E=-\varepsilon A$, with $\varepsilon>0$, can describe exponential growth in both cell radius and length. However most rod-like bacteria elongate in length while maintaining a fixed radius, suggesting a more complex shape dependence of the growth energy.

\subsection{Mechanical Energy model}
The mechanical energy of a growing cell wall is given by the sum of contributions from an internal turgor pressure, $\Pi$, acting to expand the cell volume, $V$, surface tension, $\gamma$, resisting increase in the cell surface area, $A$, and the mechanical energy of interaction with cytoskeletal bundles, $E_\text{cyto}$, which controls cell shape. That is,
\begin{equation}
E^{m}=-\Pi V+ \int dA\ \gamma + E_\text{cyto}\;.
\end{equation}
The surface tension is determined by the stored elastic energy per unit area of the cell wall, possibly offset by favorable peptidoglycan interactions at the surface~\cite{jiang2010}. Contributions to $E_\text{cyto}$ arise from MreB bundles that control cell width~\cite{jones2001,figge2004}, FtsZ filaments that drive cell wall constriction~\cite{erickson1996}, and crescentin bundles that control curvature in \textit{C.\ crescentus} cells~\cite{ausmees2003}. 

In the following Results section, we use eqn~\eqref{eqmotion1} to study shape dynamics in spherical, ellipsoidal, rod-like, and curved bacteria, by considering specific forms for the energy function $E^m$. We compare our predictions and results against available experimental observations and data.

\section{Results}
\subsection{Growth in round cells}

We first consider the simplest case of a spherical cell as a model for round bacteria like \textit{S. pneumoniae} or \textit{S. aureus}, where cytoskeletal bundles such as MreB are known to be absent ($E_\text{cyto}=0$). We model the active growth energy as $E^{a}=-\Pi_a V$, where $\Pi_a$ is the energy released per unit volume of cell wall synthesis. Neglecting cell division, the internal energy is simply given by, $E_\text{round}=-PV + \gamma A$, where $P=\Pi+\Pi_a$ is the effective growth pressure. The dynamics of the cell radius, $r$, follow from eqn~\eqref{eqmotion1}:
\begin{equation}
\frac{dr}{dt}=4\pi \mu_r r\left(Pr-2\gamma-r\frac{d\gamma}{dr}\right)\;,
\end{equation}
where $\mu_r=1/4\pi h\eta_r$ is the mobility coefficient. 
We consider two distinct models for cell wall mechanics.
If the cell wall deforms like a plastic material, $\gamma$ is a constant~\cite{boudaoud2003} such that there exists a critical radius $r_c=2\gamma/P$ (given by Laplace's law~\cite{de2004}), at which the cell size is stationary. By minimizing the energy one finds that the cell grows for $r>r_c$  and shrinks for $r<r_c$. Thus, a newborn cell must at least attain a critical size $r_c$ for growth and survival. If, however, the surface tension originates from the elastic strain energy stored in the pressurized spherical shell, we get $\gamma=\gamma_0(r/r_c)^2/2$, where $\gamma_0=Yh/2(1-\nu)$, $Y$ is the Young's modulus and $\nu$ is the Poisson ratio of the cell wall~\cite{bower2011}. In this case, the cell radius attains the steady-state value, $r_c=2\gamma_0/P$, which corresponds to an absolute minimum in the internal energy. The latter case is relevant for spherical bacteria that maintain a stable cell size before the onset of cell division~\cite{kuru2012}. Thus, a plastic cell wall can support indefinite growth if nutrient availability is optimal and division is inhibited, whereas an elastic cell wall cannot support growth beyond a threshold size $r_c$.
\begin{figure}
\centering
\includegraphics[width=\columnwidth]{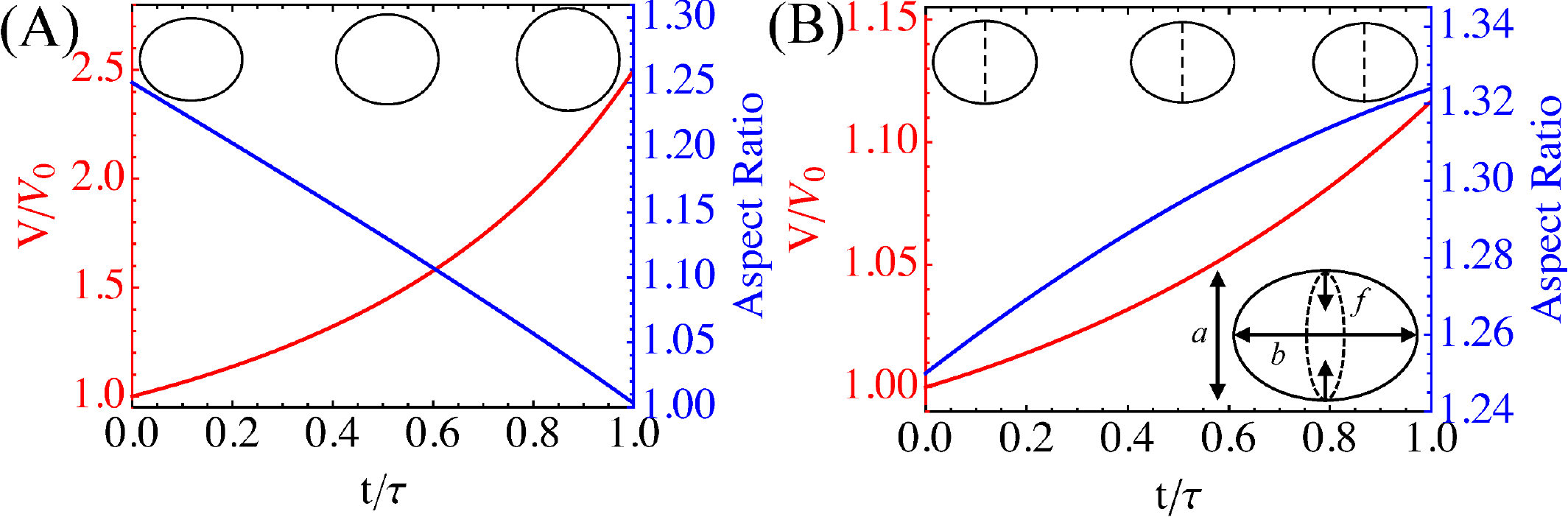}
\caption{Growth modes in ellipsoidal bacteria. (A) Oblate growth in spheroidal cells, in the absence of septum tension ($f=0$). Cell volume (red curve) and aspect ratio (blue curve), respectively, increase and decrease with time. $\tau$ defines the generation time and $V_0$ is the initial cell volume. Inset: Cell shape evolution during growth. (B) Prolate growth in spheroidal cells in the presence of tension in the septal ring, $f=0.5$. Both volume and aspect ratio grow during the cell cycle. Inset: (Top) cell shape evolution during growth. (Bottom) Schematic of a spheroidal cell with semi axes $a$ and $b$, and $f$ is the tension in the septal ring in units of $P a(t=0)^2$. Parameters: $\gamma=0.4$, $P=1$, $\mu_a=\mu_b=1.25$.}
\label{fig:round}
\end{figure}

However, in reality bacteria such as \textit{S. aureus} are not perfectly spherical but have an ellipsoidal shape. Recent experiments show that \textit{S. aureus} cells grow in volume throughout their cell cycle while their aspect ratio initially decreases followed by a period of increase~\cite{zhou2015}. For a more realistic description of \textit{S. aureus} geometry and to facilitate closer comparisons with experiments we model the shape of a \textit{S. aureus} bacterium as a spheroid, with semi-axes $a$ and $b$ ($b>a$) defining the shape parameters (Fig.~\ref{fig:round}B, inset). Their dynamics are given by
\begin{subequations}
\begin{gather}
\frac{1}{a}\frac{da}{dt}=-\frac{a\mu_a}{A}\frac{\partial E_\text{round}}{\partial a}\;,\\
\frac{1}{b}\frac{db}{dt}=-\frac{b\mu_b}{A}\frac{\partial E_\text{round}}{\partial b}\;,
\end{gather}
\end{subequations}
where $A$ and $E_\text{round}$ are the surface area and the energy of the spheroidal bacterium with $\mu_a$ and $\mu_b$ defining the growth mobility along the semi-axes $a$ and $b$ respectively. We model the active energy as $E^a=-\Pi_a V + 2\pi a f$, where $f$ is the tension due to the division septum at the midcell (Fig.~\ref{fig:round}B, inset). Net energy is thus given by $E_\text{round}=-PV + \gamma A + 2\pi a f$. Prior to the formation of the division septum, $f=0$, and the cell exhibits oblate growth such that it increases in volume but decreases in aspect ratio. As shown in Fig.~\ref{fig:round}A, an initial spheroidal cell with $b>a$ will assume a spherical shape with $b=a$. In contrast, for non-zero $f$, the cell exhibits prolate growth such that volumetric growth is accompanied by increasing aspect ratio (Fig.~\ref{fig:round}B), in agreement with recent experiments on \textit{S. aureus}~\cite{zhou2015}.

\subsection{Growth and shape control in rod-like cells}

\begin{figure}[h]
\centering
\includegraphics[width=\columnwidth]{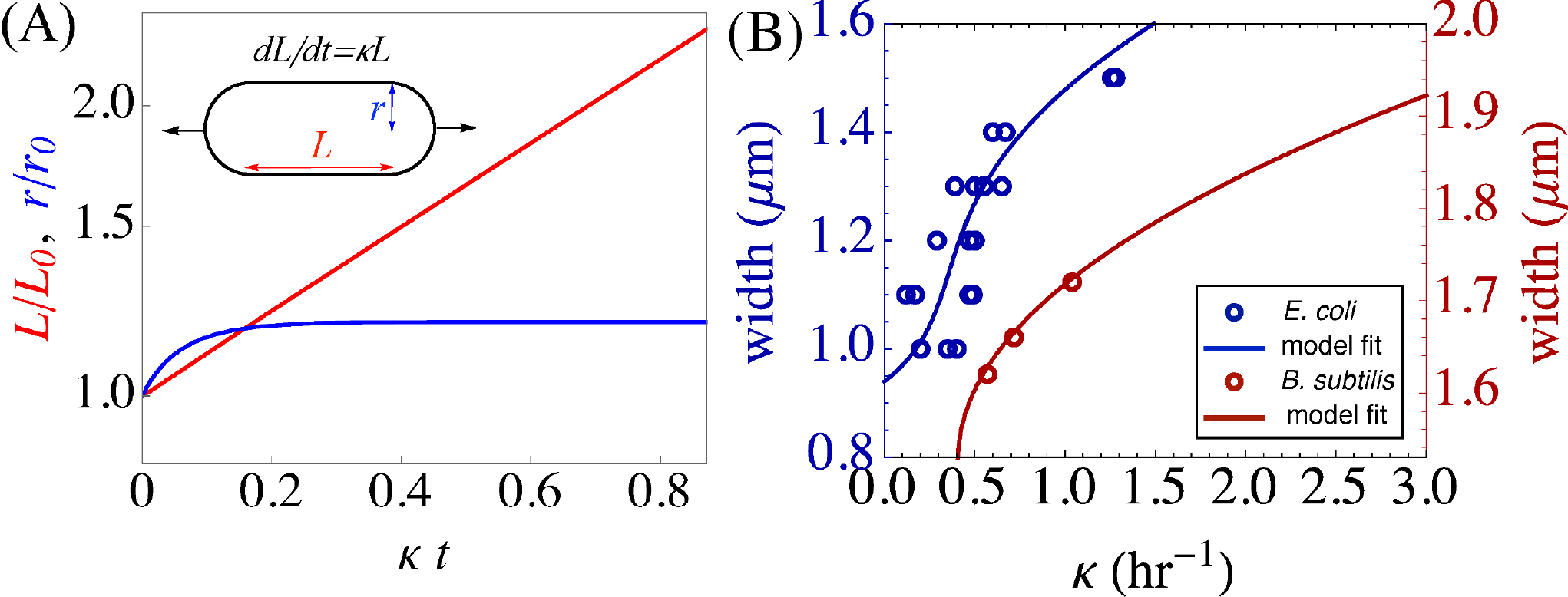}
\includegraphics[width=\columnwidth]{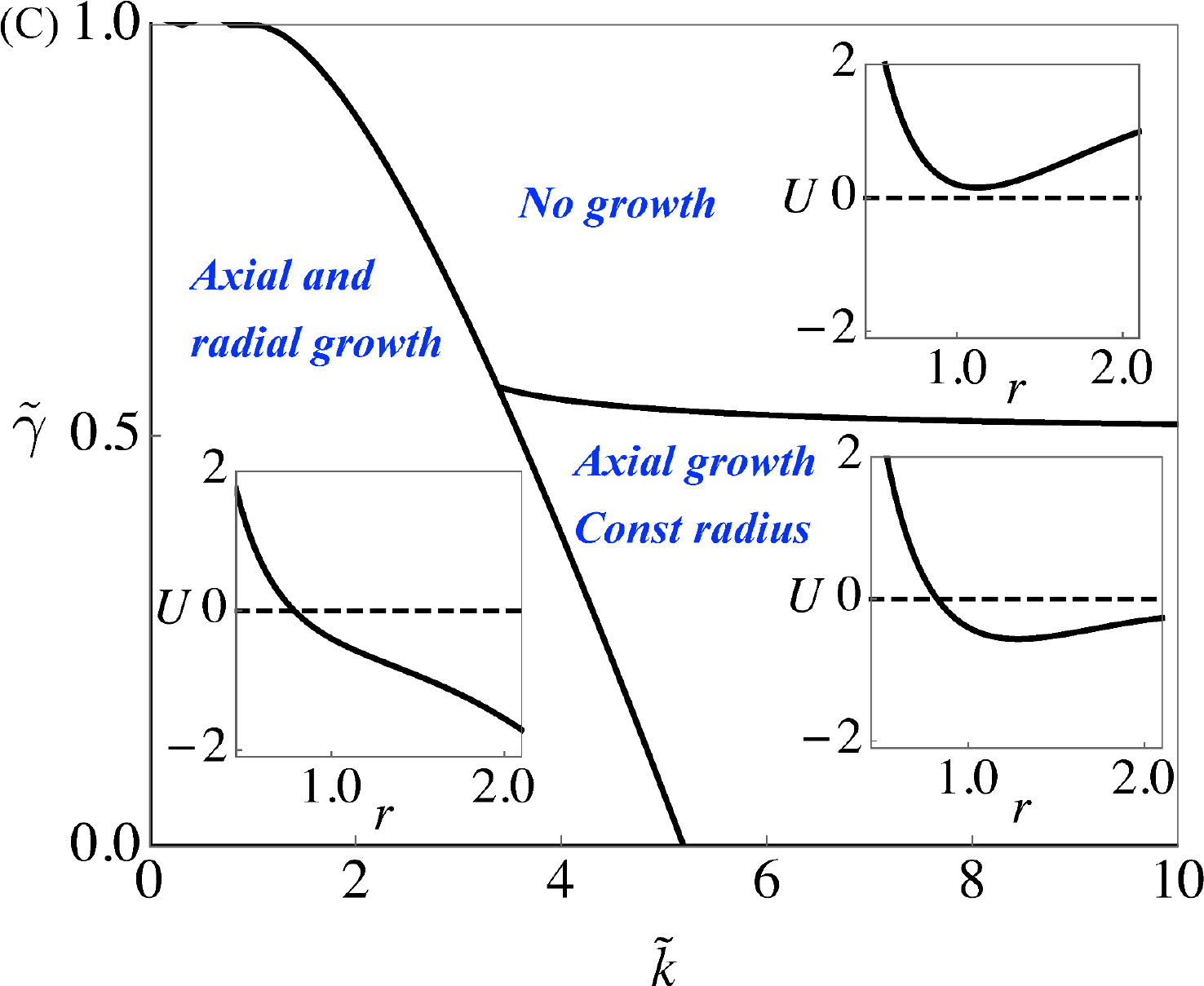}
\caption{Growth dynamics of rod-like cells. (A) Dynamics of length ($L$) and radius ($r$) normalized by their initial values ($L_0$ and $r_0$) in semi-log scale with time normalized by the timescale for growth, $\kappa^{-1}$ (see text). The surface tension and bending rigidity are $\tilde{\gamma}=0.3$ and $\tilde{k}=6$, respectively. Inset: Schematic of a longitudinally growing cylindrical cell. (B) Dependence of cell width ($2r$) on growth rate ($\kappa$). Open circles represent experimental data for {\it E. coli}~\cite{volkmer2011} (blue) and {\it B. subtilis}~\cite{sharpe1998} (red) and solid curves represent model fits. Fitting parameters: ({\it E. coli}) $\tilde{\gamma}$=0.56, $\tilde{k}$=3.2; ({\it B. subtilis}) $\tilde{\gamma}$=0.53, $\tilde{k}$=3.6. (C) Phase diagram in $\tilde{\gamma}$-$\tilde{k}$ plane showing different regions of steady-state behavior. Insets: Representative plots of energy density $U$ as a function of cell radius $r$ in the three regions of parameter space.}
\label{fig:ecoli}
\end{figure}

Rod-like cells such as \textit{E.\ coli} assume the shape of a sphero-cylinder parametrized by the radius ($r$) and the length ($L$) (Fig.~\ref{fig:ecoli}A, inset). \textit{E.\ coli} cells grow by lateral insertion of peptidoglycan material~\cite{holtje1998}. We neglect the hemispherical poles that are mechanically rigid and inert~\cite{thwaites1991}. The internal energy for the cylindrical cell is given by
\begin{equation}
E_\text{rod}=-P(\pi r^2 L) + \gamma (2\pi rL) + E_\text{cyto}\;,
\end{equation}
where $E_\text{cyto}=E_\text{width}$ is the mechanical energy for maintaining the cell width. 
$$E_\text{width}=\frac{k}{2}\int dA\ \left(\frac{1}{r}-\frac{1}{R_0}\right)^2=k\pi r L\left(\frac{1}{r}-\frac{1}{R_0}\right)^2\;,$$ where $R_0$ is the preferred radius of cross section of the cell wall and $k$ is the circumferential bending rigidity. Contributions to $k$ can come from the elasticity of glycan strands in the peptidoglycan cell wall as well as from membrane bound cytoskeletal proteins such as MreB, MreC and RodZ that are known to be responsible for maintaining rod-like cell shape~\cite{wachi1987,iwai2002,jiang2011}.

In this section, we neglect constriction to establish the basic growth dynamics.  Such a situation can be realized experimentally by suppressing division, which gives rise to filamentous cells~\cite{amir2014}. The internal energy assumes the scaling form $E_\text{rod}(r,L)=U(r)L$, where the energy density, $U$, is solely a function of the cell radius. According to eqn~\eqref{eqmotion1}, the length and the radius evolve as
\begin{subequations}
\begin{gather}
\frac{1}{L}\frac{dL}{dt}=-\mu_L\frac{U}{r}\; \label{eq:L}\\
\frac{1}{r}\frac{dr}{dt}= -\mu_r\frac{dU}{dr}\;, 
\end{gather}
\end{subequations}
where $\mu_L=1/2\pi h \eta_L$ and $\mu_r=1/2\pi h \eta_r$ are the longitudinal and radial mobility coefficients, respectively, and $\eta_L$ and $\eta_r$ are the associated viscosities. From eqn~\eqref{eq:L} the cell length grows exponentially if $U$ has a minimum at $r=r_s$ such that $U(r_s)<0$ and is a constant. Fig.~\ref{fig:ecoli}A shows the dynamics of length and radius in the regime of parameters that allow exponential elongation at constant radius. 

Assuming that the cell maintains a constant effective growth pressure, we can rescale the energy density by $U_0=\pi P R_0^2$. The shape dynamics are then controlled by two dimensionless parameters, $\tilde{\gamma}=\gamma/P R_0$ and $\tilde{k}=k/PR_0^3$. In the limit $\tilde{k}\gg 1$, the cell radius approaches $R_0$, and the cell assumes a stationary shape defined by the value $\tilde{\gamma}= 1$. As such, the numerical values for $\tilde{\gamma}$ and $\tilde{\kappa}$ are cell type dependent (Table 1). For Gram-negative \textit{E. coli} cells with Young's modulus $Y\simeq 25-50$ MPa, $h\simeq 3$ nm and $P\simeq 0.3$ MPa~\cite{deng2011}, the estimated values for $\tilde{\gamma}$ lie in the range $0.25-0.5$. Whereas for Gram-positive \textit{B. subtilis} cells with smaller values for Young's modulus $Y\simeq 15-30$ MPa, and larger values for thickness and pressure, $h\simeq 30-40$ nm and $P\simeq 1.5$ MPa~\cite{lan2007}, the estimated values for $\tilde{\gamma}$ lie in the range $0.3-0.8$. In Fig.~\ref{fig:ecoli}B we show the dependence of the cell width ($2r$) on the rate of exponential growth, $\kappa=-\mu_L U(r)/r$, for parameter values corresponding to \textit{E. coli} and {\it B. subtilis}. The parameters are determined by fitting our model prediction to the available data on \textit{E. coli}~\cite{volkmer2011} and {\it B. subtilis}~\cite{sharpe1998}. In agreement with experimental data~\cite{nanninga1988,sharpe1998,volkmer2011,taheri2015}, our model quantitatively captures the positive correlation between $\kappa$ and $r$ for both cell types. The predicted cell width for {\it E. coli} is more sensitive to changes in growth rate, presumably due to the fact that the cell wall is softer and thinner in {\it E. coli} because it is gram negative.

The steady-state behavior at different values of $\tilde{\gamma}$ and $\tilde{k}$ is shown in Fig.~\ref{fig:ecoli}C. The corresponding plots of the energy densities are shown in the insets to Fig.~\ref{fig:ecoli}C. While radial growth occurs for smaller values of $\tilde{k}$, exponential elongation with constant radius occurs for $\tilde{\gamma} \lesssim 0.5$ and $\tilde{k} \gtrsim 4$. Using the typical range of estimates for the internal pressure in gram-negative bacteria $P\simeq 0.1$-0.5 MPa~\cite{thwaites1991,deng2011} and the preferred radius of cross-section $R_0\simeq 0.1$-0.5 $\mu$m~\cite{jiang2011}, we predict the upper bound on surface tension to be $\gamma_\text{max}\simeq 50$ nN/$\mu$m and a lower bound on the circumferential rigidity to be $k_\text{min}\simeq0.4$ nN$\mu$m. These values are in agreement with estimates based on mechanical measurements~\cite{wang2010b,deng2011,wright2014} and suggest that rod-like bacteria operate close to the triple point in Fig.~\ref{fig:ecoli}C.

\subsection{Response to shape perturbations}
\begin{figure}
\centering
\includegraphics[width=\columnwidth]{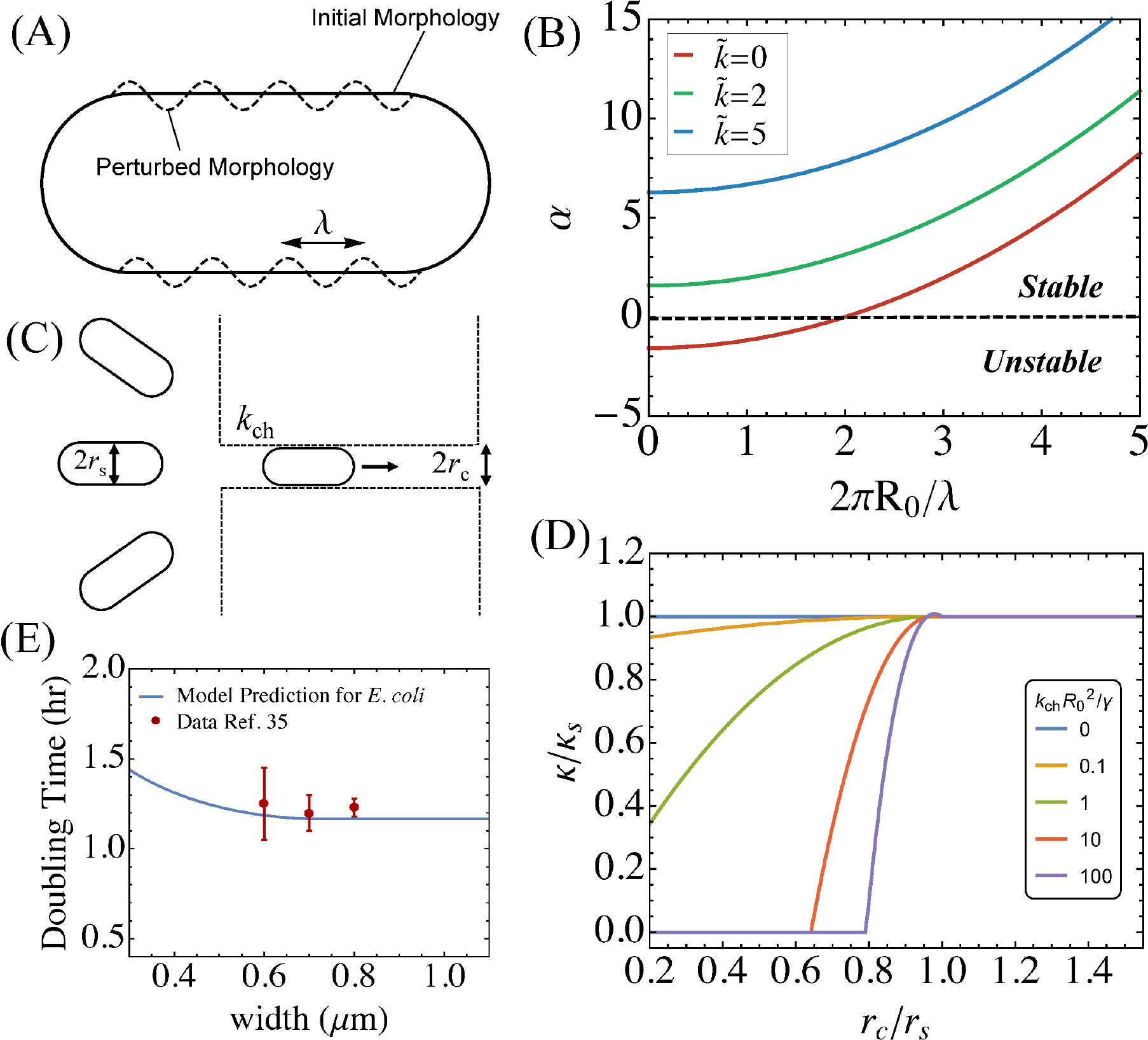}
\caption{Response of growth dynamics to shape perturbations. (A) An initial cylindrical cell (solid line) undergoing morphological perturbation of harmonic form (dashed curve). (B) Decay rate $\alpha$ for the amplitude of the harmonic perturbation of the cell radius, as a function of the dimensionless wavenumber $2\pi R_0/\lambda$ at various values of $\tilde{k}$. The system is stable to perturbations for $\alpha>0$ and unstable for $\alpha<0$. (C) Schematic of rod-like bacteria squeezed into narrow channels of width $2r_c$ and rigidity $k_{ch}$~\cite{mannik2009}. When not constrained to grow in the microchannels, the cells grow while maintaining a constant diameter $2r_s$. (D) Dependence of longitudinal growth rate (normalized by the steady-state value $\kappa_s$) on the channel radius (normalized by steady-state cell radius $r_s$) for different values of the dimensionless parameter $k_{ch}R_0/\gamma$, describing the relative stiffness of the channel to the cell wall. Parameters: $\tilde{\gamma}=0.45$, $\tilde{k}=6$. (E) Doubling time vs channel width for {\it E. coli} cell parameters (Table 1). Solid blue curve is the model prediction. The data (red) are taken from Ref.~\cite{mannik2009}.}
\label{fig:pert}
\end{figure}

\begin{table*}[t]
\small
  \caption{\ List of parameters used in the energy model}
 \label{table:params}
  \begin{tabular*}{\textwidth}{@{\extracolsep{\fill}}lllll}
  \hline
  Parameter & Description & Function & Associated Molecules & Numerical Estimate \\
  \hline
$\Pi$ & Turgor Pressure & Cell wall expansion & Peptidoglycan & 0.3 MPa ({\it E. coli})\\
$\ $ &  &  &  & 1.5 MPa ({\it B. subtilis})~\cite{lan2007,deng2011}  \\
$\Pi_a$ & Growth pressure & Cell wall synthesis & PBPs, MreB & 0.4 MPa ({\it E. coli})\\
$\ $ &  &  &  & 1.5 MPa ({\it B. subtilis})  \\
%\hline
$\gamma$ & Surface tension & Cell shape maintenance & Peptidoglycan & 19 nN/$\mu$m ({\it E. coli})   \\
$\ $ &  &  &  & 113 nN/$\mu$m ({\it B. subtilis})~\cite{lan2007,deng2011} \\
%\hline
$k$& Circumferential bending rigidity & Cell width control & MreB, MreC, RodZ & 0.03 MPa$\mu$m$^3$ ({\it E. coli})\\
$\ $ &  &  &  &  0.3 MPa$\mu$m$^3$ ({\it B. subtilis})~\cite{lan2007}   \\
%\hline
$R_0$& Preferred radius of cross-section & Cell width maintenance & MreB, MreC, RodZ & 0.38 $\mu$m ({\it E. coli}) \\
$\ $ &  &  &  &  0.43 $\mu$m ({\it B. subtilis})~\cite{jiang2011,mannik2009}\\
%\hline
$k_c$ & Longitudinal bending rigidity & Cell curvature control & Crescentin & 1.5 nN$\mu$m$^2$~\cite{jiang2012,wright2014}  \\
%\hline
$R_c$ & Preferred radius of curvature & Cell curvature maintenance & Crescentin & 2-6 $\mu$m \\
%\hline
$\varepsilon$ & Chemical potential for growth & Septum synthesis & PBPs, divisomes & $>$12 nN/$\mu$m (prediction) \\
%\hline
$f$ & Line tension & Constriction force & FtsZ & 8-80 pN ~\cite{lan2007,allard2009}\\ 
\hline
\end{tabular*}
\end{table*}
Having discussed growth and shape dynamics under steady environmental conditions, we now consider how rod-like cells modulate their growth dynamics in response to morphological perturbations. 
It has been experimentally observed that upon addition of A22, which causes disassembly of MreB, wave-like bulges form on the cell wall~\cite{jiang2011}. Thus, the loss of MreB, which corresponds to lower values of $\tilde{k}$, can induce morphological instabilities in the cell wall. Motivated by this observation, we now investigate the robustness of the rod-like geometry to external perturbations as a function of $\tilde{k}$. We examine the stability of a cylindrical shape under a small periodic perturbation of the steady-state cell radius $r_s$ along the axial direction $z$: $r(z,t)=r_s(t) + \delta r(t) \cos{(2\pi z/\lambda)}$, where $\delta r\ll r_s$ is the amplitude and $\lambda$ is the wavelength of the perturbation (Fig.~\ref{fig:pert}A). To leading order in $\delta r$, the internal energy of the cell integrated over one cycle of the perturbation is given by $E_\text{rod}/\lambda=U + \alpha(\lambda) \delta r^2 + \mathcal{O}(\delta r^4)$, where the decay rate of the perturbation, $\alpha(\lambda)$, is an increasing function of the wavenumber $2\pi/\lambda$ for all values of $\tilde{k}$ (Fig.~\ref{fig:pert}B). The stability of the cylindrical shape is determined by the sign of $\alpha$. For $\alpha<0$ the cylindrical shape is unstable to perturbations of wavelength greater than $\lambda$, such that wave-like bulges nucleate on the cell-wall with growing amplitude. For $\tilde{k}=0$ the cylindrical shape is unstable for $\lambda>\lambda_\text{min}=2\pi\sqrt{\gamma r_s/P}$. However, as $\tilde{k}$ increases beyond a critical value, the cylindrical shape is stable to perturbations of any wavelength. 

Another experiment that allows examining predictions of our model involves studying bacterial growth and movement in sub-micron microfluidic channels that geometrically confine growth~\cite{mannik2009,mannik2012} (Fig.~\ref{fig:pert}C). Rod-like bacteria such as {\it E. coli} or {\it B. subtilis} are able to grow in very narrow microfluidic channels of width comparable to or even smaller than their unperturbed diameters. Furthermore, bacterial cell walls can deform (the ceiling of) the microchannels, which are made of elastic material such as PDMS. To verify if our growth model can capture the experimental results~\cite{mannik2009}, we include the elastic interaction between the channel and the cell wall as $E_\text{int}(r_c<r_s)=\frac{1}{2}k_{ch}\int dl (r-r_c)^2$, where $k_{ch}$ is the channel stiffness, $r$ is the radius of the bacterial cell wall, $r_c$ is the radius of the cylindrical channel and $r_s$ is the steady state radius of a freely growing bacterium. If the channel is wider than $r_s$ then there is no interaction between the channel and the cell wall, $E_\text{int}(r_c>r_s)=0$. The total energy is then given by $E_\text{rod}+E_\text{int}$. By solving the coupled equations for cell length and radius (eqn \eqref{eqmotion1}) we derive the growth rate dependence on channel width. For channels wider than unperturbed cell radius, $r_c>r_s$, the cells grow at a constant rate $\kappa=\kappa_s$. For $r_c<r_s$, the dependence of $\kappa$ on $r_c$ is controlled by the dimensionless parameter $k_{ch} R_0/\gamma$, describing the stiffness of the channel relative to the cell wall (Fig.~\ref{fig:pert}D). For channels softer than the cell wall, we find that the growth rate is insensitive to channel radius. However if the channel is stiffer than the cell wall, we predict that the growth rate increases monotonically with $r_c$, and no growth occurs below a critical channel radius. We quantitatively compare our model predictions with the experimental data on doubling times of {\it E. coli} cells vs channel radius~\cite{mannik2009}. As shown in Fig.~\ref{fig:pert}E, our model is in good quantitative agreement with the data and predicts that the longitudinal growth rate is insensitive to changes in channel width beyond 0.5 $\mu$m.

\subsection{Curved cells}

\begin{figure}
\centering
\includegraphics[width=\columnwidth]{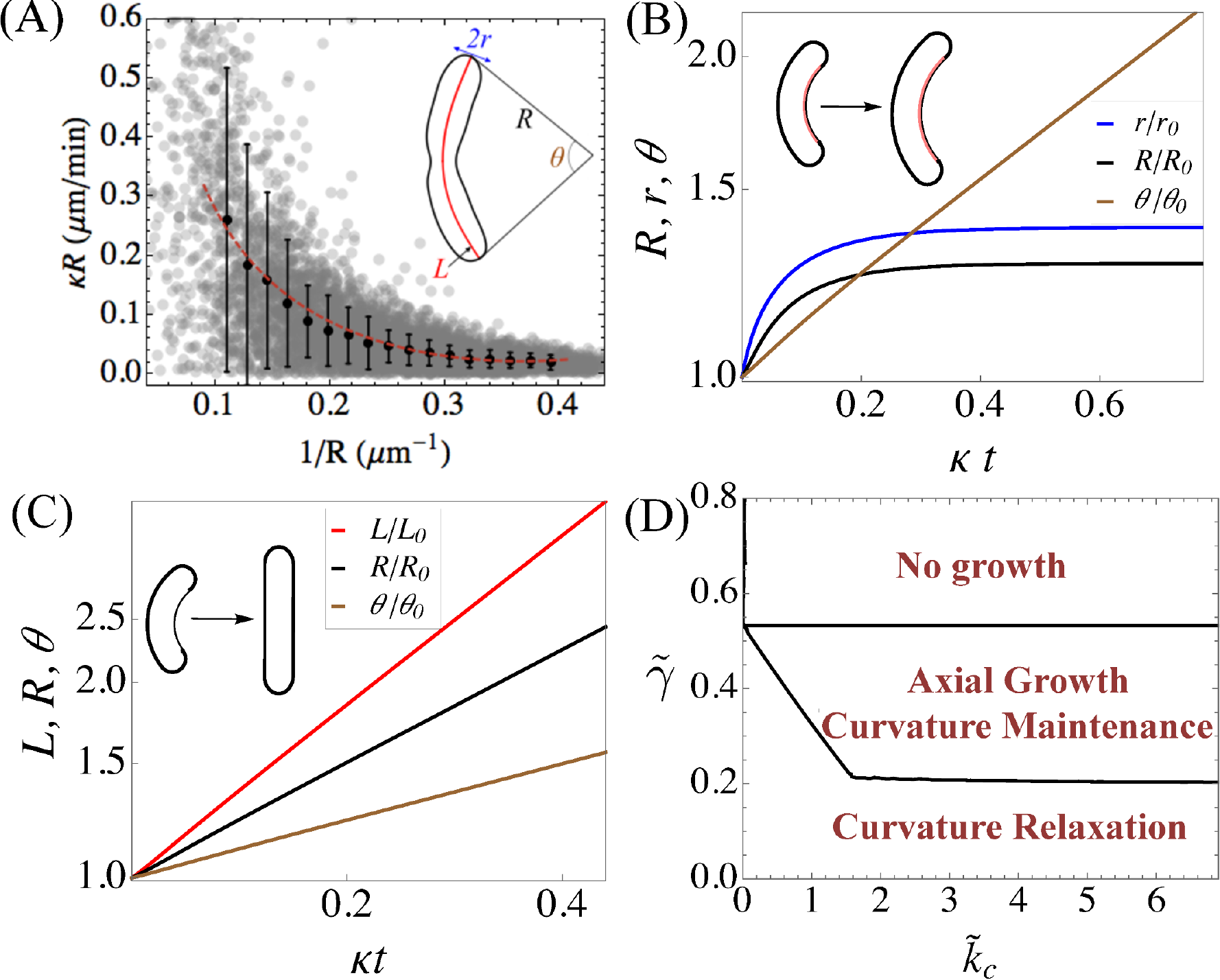}
\caption{Growth and shape dynamics of curved cells. (A) Inset: Schematic of a \textit{C.\ crescentus} cell contour and the shape parameters~\cite{wright2014}. Dependence of angular growth rate ($\kappa R$) on the curvature of the cell $1/R$. We determine the cell curvature and the spanning angle for each generation from the splined contour of the cell boundary. We then obtain the angular growth rate, $\kappa$, by fitting an exponential to the data for $\theta(t)$. Gray points indicate single-generation data. Experimental binned data~\cite{wright2014} are shown by solid black circles and the model fit is given by the red dashed curve. Error bars represent $\pm$1 standard deviation. (B) Curvature maintenance in the presence of crescentin at a value of the dimensionless bending rigidity $\tilde{k}_c=k_c/PR_0^4=4$. Dynamics of the radius of curvature ($R$), radius of cross-section ($r$), and spanning angle ($\theta$) normalized by their initial values in a semi-log plot. (C) Cell straightening in the absence of crescentin ($k_c=0$). Cell length ($L$), radius of curvature ($R$) and spanning angle ($\theta$) grow exponentially (shown in a semi-log plot, normalized by the respective initial values). Parameters: $\tilde{k}=5$, $\tilde{\gamma}=0.3$, $R_c=R_0$. (D) Phase diagram in $\tilde{\gamma}$-$\tilde{k}_c$ plane illustrating the steady-state growth behaviors.}
\label{fig:cc}
\end{figure}

As an example of a curved cell, we explore the shape dynamics of a \textit{C.\ crescentus} bacterium. Note that the results are not specific to crescent-shaped bacteria, and apply equally well to helical bacteria, for example. We model the geometry of a \textit{C.\ crescentus} cell by a toroidal segment parametrized by the radius of cross section $r$, centerline radius of curvature $R$, and the spanning angle $\theta$ (Fig.~\ref{fig:cc}A, inset). Experiments have shown that the curvature of \textit{C.\ crescentus} cells is maintained by intermediate filament-like bundles of crescentin proteins that adhere to the concave face of the cell wall~\cite{cabeen2009}. Although the molecular mechanism by which crescentin maintains cell curvature is not precisely known, proposed models include modulation of elongation rates across the cell wall~\cite{cabeen2009}, which can originate from  bundling with a preferred curvature~\cite{jiang2012}. We thus model the curvature energy in the crescentin bundle as
$$E_\text{cres}=\frac{k_c}{2}\int_0^{\ell_c} d\ell \left(C-\frac{1}{R_c}\right)^2\;,$$
where $\ell_c=(R-r)\theta$ is the contour length of the crescentin bundle, $C$ is the longitudinal cell wall curvature, $k_c$ is the bending rigidity, and $R_c$ is the intrinsic radius of curvature of the bundle. The energy term, $E_\text{cres}$, accounts for the compressive stresses generated by the crescentin bundle on one side of the cell wall, thereby leading to differential growth across the sidewall. The total internal energy is given by 

\begin{equation}\label{eq:encurve}
E_\text{curv}=-PV + \gamma A + E_\text{cyto}\;,
\end{equation}
where $E_\text{cyto}=E_\text{width} + E_\text{cres}$. In the presence of crescentin, $k_c\neq 0$, the internal energy has the scaling form $E_\text{curv}(R,r,\theta)=\theta U_c(r,R)$. The dynamics of the shape parameters $R$, $r$, and $\theta$ follow from eqn~\eqref{eqmotion1}, characterized by the viscosity parameters $\eta_R$, $\eta_r$ and $\eta_\theta$, respectively. The cell exhibits \textit{hoop-like growth}~\cite{mukhopadhyay2009} with $R$ and $r$ remaining constant and $\theta$ growing exponentially as
\begin{equation}\label{eq:theta}
\frac{1}{\theta}\frac{d\theta}{dt}=-\mu_\theta \frac{U_c(r,R)}{rR}\;,
\end{equation}
with a rate $\kappa=-\mu_\theta U_c/rR$, where $\mu_\theta=1/(2\pi h\eta_\theta)$ is the hoop growth mobility. The shape variables $R$ and $r$ attain constant steady-state values determined by the global minimum of $U_c(r,R)$ (Fig.~\ref{fig:cc}B). Our model predicts that the angular growth rate, $\kappa R$, is a decreasing function of the curvature, $1/R$ (Fig.~\ref{fig:cc}A, red curve). This coupling between angular dynamics and curvature arises through the curvature dependence of the bending energy ($E_\text{cres}$) that increases with cell curvature. The growth rate is proportional to $-E_\text{cres}$ through the dependence of $U_c$ on $R$, and $\kappa$ is consequently larger for straight cells (larger R). 

We test this prediction of our model with our experimental shape data on single \textit{C. crescentus} cells~\cite{wright2014}. The scatter plot showing the dependence of angular growth speed on cell curvature demonstrates that angular growth is slower for curved cells (Fig.~\ref{fig:cc}A). The fitted model (red) is in excellent agreement with the binned data (black points) at the value of the fitting parameter $\tilde{k}_c=1.75$, which further constrains the physical values of $\tilde{\gamma}$ in the range 0.2-0.5, as discussed below (see Fig.~\ref{fig:cc}D).
%This prediction hints at a possible role of cell curvature on growth and fitness, which to the best of our knowledge has not yet been verified experimentally. This result can be quantitatively tested by measuring longitudinal growth rates of \textit{C.\ crescentus} cells treated with drugs that alter curvature or in \textit{E.\ coli} cells subjected to sustained periods of microfluidic flow \cite{amir2014}.

In the absence of crescentin ($k_c=0$), the internal energy assumes the scaling form $E_\text{curv}(R,r,\theta)=\theta R U(r)$, such that both $R$ and $\theta$ grow exponentially as expected during \textit{self-similar growth}. This leads to cell straightening (Fig.~\ref{fig:cc}C), as observed for cells lacking creS~\cite{cabeen2009}. The cell curvature, $C=1/R$, decays as $dC/dt=-\kappa' C$, whereas the angle grows according to $d\theta/dt=\kappa''\theta$, with $\kappa'/\kappa''=\eta_\theta/\eta_R$. The cell length ($L=R\theta$) consequently grows exponentially with a rate $\kappa'+\kappa''$. The ratio $\kappa'/(\kappa'+\kappa'')$, which quantifies the propensity of cell straightening, has been experimentally determined to be $\simeq0.57$~\cite{sliusarenko2010}. We thus estimate the ratio of viscosities characterizing the angular and curvature dynamics to be $\eta_\theta/\eta_R\simeq1.3$, implying that angular growth is slower than the decay of cell curvature.
The steady-state behavior at different values of $\tilde{\gamma}$ and $k_c$, which control cell size and shape respectively, is shown in Fig.~\ref{fig:cc}D for fixed values of pressure and width. In particular, we find that the cell elongates exponentially while maintaining a constant curvature in the range $0.2<\tilde{\gamma}<0.5$. At smaller values of surface tension, $\tilde{\gamma}$, the cell wall cannot support curvature-induced stresses and relaxes to a straight morphology. For $\tilde{\gamma}>0.5$, the cell maintains a stationary size and shape.

\subsection{Cell wall constriction}
\begin{figure}
\centering
\includegraphics[width=\columnwidth]{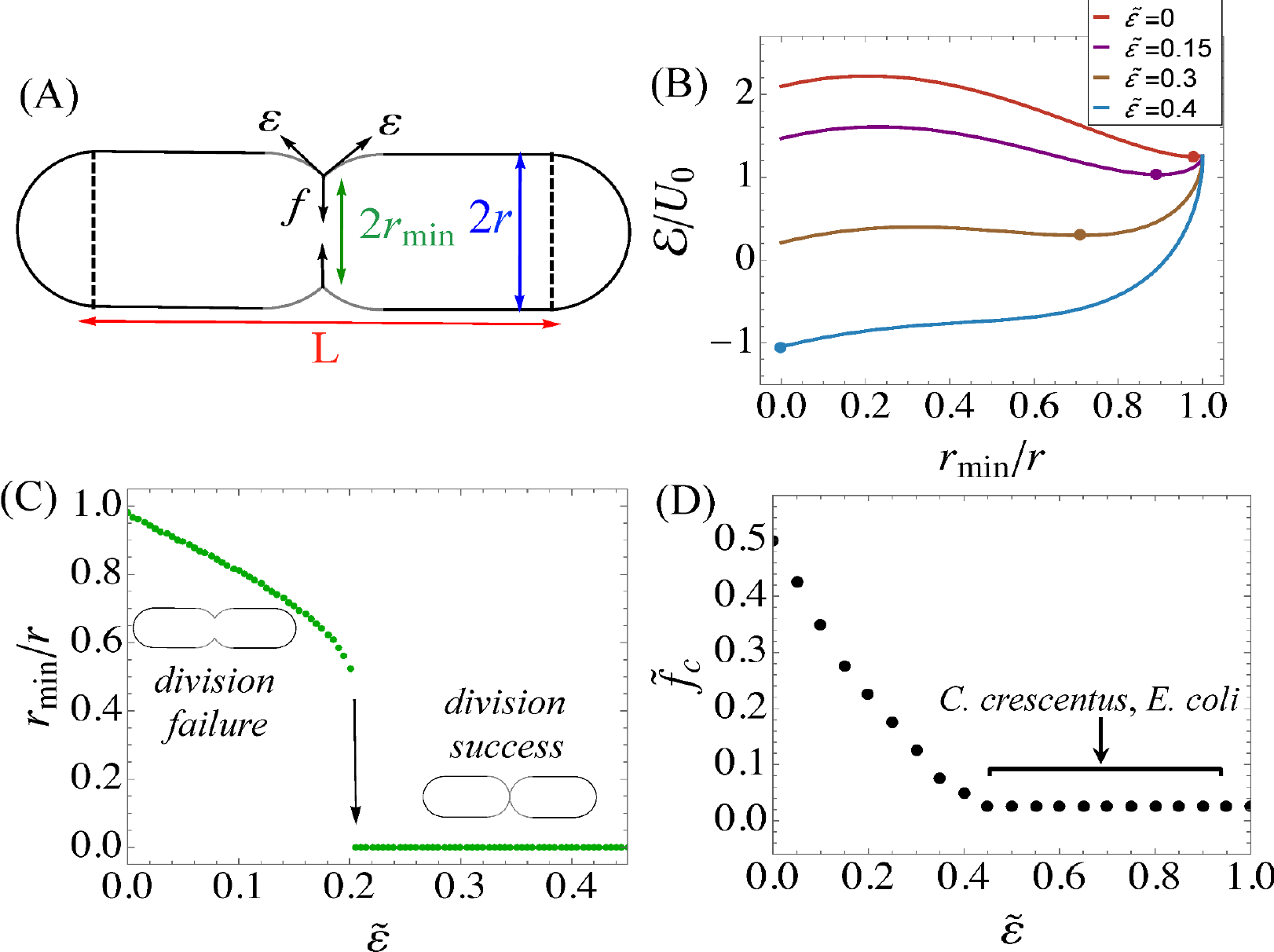}
\caption{Mechanics of cell wall constriction. (A) Schematic of the constricting cell. The arrows indicate the forces driving constriction arising from tension ($f$) and growth ($\varepsilon$). (B) Constriction energy $\mathcal{E}$ as a function of $r_\text{min}/r$ at a fixed tension $\tilde{f}=f/PR_0^2=0.2$ and different values of the dimensionless chemical potential $\tilde{\varepsilon}=\varepsilon/PR_0$: 0 (red), 0.15 (purple), 0.3 (brown) and 0.4 (blue). The corresponding minima are indicated by solid circles. (C) Bifurcation diagram showing the dependence of the division order parameter $r_\text{min}/r$ (green) on the chemical potential $\tilde{\varepsilon}$. Cell division is successful for $\tilde{\varepsilon}>\tilde{\varepsilon}_c\simeq0.2$, where a discontinuous transition occurs between partial and full constriction ($\tilde{f}=0.2$). (D) Critical tension $\tilde{f}_c$ required for full constriction as a function of the chemical potential $\tilde{\varepsilon}$. For $\tilde{\varepsilon}>0.45$, no mechanical force is required for cell division, with $\tilde{f}_c<0.015$.}
\label{fig:division}
\end{figure}

We now study how cell wall growth couples with constriction in bacteria. The onset of cell wall constriction influences the overall shape dynamics of the cell. For simplicity, we first consider the case of a rod-like bacterium. Bacterial cell division is driven by a large complex of proteins, known as the divisome, that assembles the Z-ring near the mid-plane of the cell~\cite{erickson2010}. The Z-ring comprises FtsZ filaments that form a patchy band structure~\cite{holden2014}. It is believed that these filaments generate constrictive and bending forces~\cite{erickson1996}. In addition the divisome triggers peptidoglycan synthesis and directs formation of the septum~\cite{typas2011}. We assume that the shape of the constriction zone is defined by two intersecting and partially formed hemispheres with radii $r$, equal to the radii of the new poles (Fig.~\ref{fig:division}A). The shape parameter defining the mid-cell radius, $r_\text{min}(t)$, equals $r$ at the onset of constriction and reaches 0 at the completion of division. We assume that the Z-ring proteins exert a mechanical tension $f$ on the cell wall and trigger septal growth by releasing an energy $\varepsilon$ per unit surface area. The chemical potential $\varepsilon$ is related to the activity of MreB and penicillin-binding proteins (PBPs) that synthesize peptidoglycan by localizing to the division site near the mid-plane of the cell~\cite{daniel2000,divakaruni2007}. These active mechanisms contribute an energy $E^a=f (2\pi r_\text{min}) -\varepsilon S$, where $S$ is the septal surface area given by $S=4\pi r\sqrt{r^2-r_\text{min}^2}$. With $E_\text{cyto}=E_\text{width}$, the energy of the constricting cell thus takes the scaling form $E_\text{rod}(r,L,r_\text{min})=U(r) L + \mathcal{E}(r_\text{min},r)$, where $\mathcal{E}$ defines the effective energy of constriction. Therefore, the steady-state values for $r_\text{min}$ are controlled by the tension $f$ and the chemical potential $\varepsilon$. 

To determine the minimum values of $\varepsilon$ and $f$ that are required to achieve full constriction, we first examine the dependence of $\mathcal{E}$ on $r_\text{min}$ at different values of the dimensionless chemical potential $\tilde{\varepsilon}=\varepsilon/PR_0$ while keeping $f$ fixed (Fig.~\ref{fig:division}B). At $\tilde{\varepsilon}=0$ the energy is minimized for $r_\text{min}\simeq r$, and no constriction occurs. As $\tilde{\varepsilon}$ is increased, the local minimum of the energy at $r_\text{min}/r\simeq1$ shifts towards a more constricted state, but division is still unsuccessful. For $\tilde{\varepsilon}\gtrsim \tilde{\varepsilon}_c$, the local minimum is lost in favor of a global minimum at $r_\text{min}=0$, corresponding to a fully constricted state. 

This energy minimization approach reveals the fundamental mechanism behind constriction: cell shape maintenance enforces a competition between $\varepsilon$ (and $f$) that minimizes the mid-plane perimeter and the surface tension $\gamma$ that resists the associated increase in surface area~\cite{turlier2014}. The steady-state ratio $r_\text{min}/r$ obtained by minimizing the energy functional, gives us an order parameter for cell division, such that division is unsuccessful for $\tilde{\varepsilon}<\tilde{\varepsilon}_c$ and successful for $\tilde{\varepsilon}>\tilde{\varepsilon}_c$ ($r_\text{min}=0$). The bifurcation diagram in Fig.~\ref{fig:division}C shows the dependence of the order parameter $r_\text{min}/r$ as a function of $\tilde{\varepsilon}$. For smaller values of $\tilde{\varepsilon}$, $r_\text{min}/r$ decreases, whereas for $\tilde{\varepsilon}>\tilde{\varepsilon}_c\simeq 0.2$ (when $\tilde{f}=0.2$), there is a discontinuous transition to a fully constricted state. The prediction of a threshold force for completion of constriction could be tested experimentally by treating cells with controlled amounts of Divin, a small molecule inhibitor of bacterial divisome assembly that reduces peptidoglycan remodeling and prevents cytoplasmic compartmentalization~\cite{eun2013}. Consistent with this suggestion, it was found experimentally that a threshold amount of Divin is required to inhibit bacterial cell division.

%(E) Zoomed in schematic of the division septum with active regions of growth labelled in blue. (F) Pinch-off dynamics of $w_\text{min}(t)$ and its dependence on septal growth rate $\kappa_s$ with $l_{s0}/w_\text{min}(0)=1.2$
Having discussed the mechanisms for cell constriction, it is pertinent to consider the relative contributions of the mechanical tension $f$ and the chemical potential $\varepsilon$ in executing cell wall constriction. Fig.~\ref{fig:division}D shows the dependence of the critical force $f_c$ required for full constriction on the magnitude of the chemical potential $\varepsilon$. While a large mechanical force is required for low values of $\varepsilon$, we predict that for $\tilde{\varepsilon}\gtrsim0.45$ little (or no) mechanical force is required to complete division. Using $P=0.03$ MPa~\cite{deng2011} and $R_0=0.4$ $\mu$m, we predict a numerical value for the upper bound of the Z-ring mechanical force $f_c^\text{max}\simeq72$ pN which translates to $\tilde{f}_c=0.015$ in dimensionless units. This estimate is consistent with the mechanical properties of FtsZ filament bundles~\cite{lan2009}. Previous models have also suggested that a force in the range $8$-$80$ pN (0.0017-0.017 in our dimensionless units) is sufficient for pinch-off during division of rod-like bacteria~\cite{lan2007,allard2009}. We thus claim that typical rod-like bacterial cells operate in the regime $\tilde{\varepsilon}>0.45$ ($\varepsilon>$12 nN/$\mu$m) {\it such that constriction is entirely driven by cell wall synthesis at the septum}. The predicted minimum value for the chemical potential is roughly one-fourth of the surface tension measured for Gram-negative bacteria~\cite{deng2011}.

\section{Conclusions}
\begin{figure}
\centering
\includegraphics[width=0.8\columnwidth]{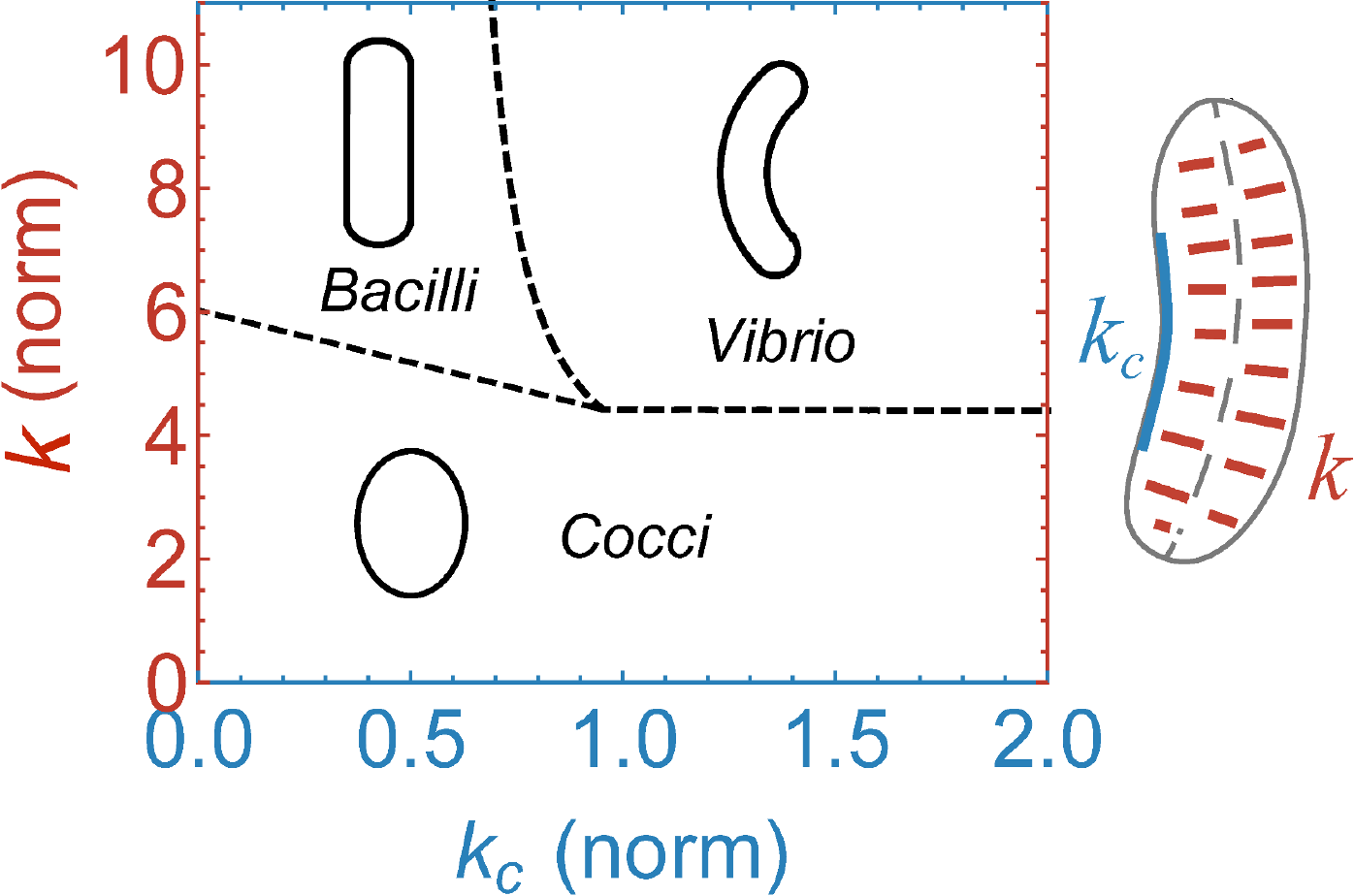}
\caption{Bacterial polymorphism. Shape stability diagram as functions of longitudinal rigidity, $k_c$ (normalized by $PR_0^4$), and circumferential rigidity, $k$ (normalized by $PR_0^3$).}
\label{fig:shapes}
\end{figure}
How cells regulate their shapes and sizes through the processes of growth and division poses a fundamental question at the interface of physics and biology. To address this fundamental question, we have developed a broadly applicable model for the shape dynamics of growing cell walls that are driven by mechanical and active forces (eqn~\eqref{eqmotion1}). Our model takes advantage of recent technological advances in single cell imaging~\cite{wang2010,iyer-biswas2014,campos2014,wright2014} that have yielded unprecedented amounts of quantitative information about the shapes of single bacteria as they grow and divide. The active forces arise from proteins driving cell wall growth and constriction, whereas the mechanical forces arise from tensions in the peptidoglycan cell wall and associated cytoskeletal bundles. The equations for the shape dynamics in combination with the appropriate energy models (see Table~\ref{table:params} for a summary of the model parameters), describe a wide range of phenomena that occur in bacterial cells, including exponential growth, steady-state sizes, shape robustness and constriction. Using the energy model, we demonstrate how width and curvature control can be achieved in bacterial cells and discuss the mechanical instabilities that can lead to morphological transformations (Figs.~\ref{fig:pert} A,B and~\ref{fig:cc}C). In Fig.~\ref{fig:shapes} we show the shape stability diagram for our energy model as functions of the mechanical parameters controlling longitudinal and circumferential curvatures of the cell wall. Our model can reproduce different families of known bacterial shapes (cocci, bacilli, vibrio) by varying the mechanical rigidities controlling the curvatures of the cell wall.

In this paper we obtained the following key conclusions:
\begin{itemize}
\item Exponential growth in cell size requires a constant amount of energy dissipation per unit volume.
\item Cell shape, as opposed to simply size, controls the rate of exponential growth in cell size.
\item Cell division can be explained as a discontinuous (first-order) shape transformation controlled by the interplay between cell wall surface tension and the chemical potential required for the addition of new cell wall material.
\item Cell growth and constriction are both driven by the addition of new cell wall material, and thus their kinetics are same. This insight provides a physical explanation for the recent experimental observation that a single time scale governs growth and division~\cite{iyer-biswas2014}.
\end{itemize}

The microscopic formulation of the equations of motion makes it convenient for their adoption in computational modeling of cell wall growth and morphology. It is, however, important to recognize that the underlying structure of the bacterial cell wall is highly dynamic, and cellular mechanical properties may fluctuate due to molecular scale defects and stochastic forces. Our model is thus valid on timescales comparable to measurable cell wall growth ($\sim$minutes) that are much larger than the timescales of molecular processes ($\sim$seconds) involving peptidoglycan bond rupture and subsequent insertions of new cell wall material. In future work we aim to incorporate the effects of stochasticity and spatiotemporal variations in cellular material parameters to better understand the statistical mechanics of shape fluctuations in living cells.

\section*{Acknowledgements}
We thank Charles Wright and Srividya Iyer-Biswas for help with reproducing the experimental data shown in Fig.~\ref{fig:cc}A from ref.~\cite{wright2014}. We gratefully acknowledge funding from the NSF Physics of Living Systems program (NSF PHY-1305542), NSF Materials Research Science and Engineering Center (MRSEC) at the University of Chicago (NSF DMR-1420709). NFS acknowledges partial support from the ONR NSSEFF program. We thank an anonymous reviewer for bringing Refs. 29, 30, 37, 38 to our attention.


%\balance


%%%REFERENCES%%%
%\bibliography{ref} %You need to replace "rsc" on this line with the name of your .bib file
%\bibliographystyle{rsc} %the RSC's .bst file
%

\end{document}